\documentclass[12pt, a4paper]{article}
\input epsf
\newcommand{\sect}[1]{ \section{#1} }

\newcommand{\ve}{\left( \begin{array}{r}}
\newcommand{\ev}{\end{array} \right)}
\newcommand{\ar}{\left( \begin{array}{rr}}
\newcommand{\ra}{\end{array} \right)}
\newcommand{\arr}{\left( \begin{array}{rrrr}}
\newcommand{\arrr}{\left( \begin{array}{rrrrrr}}
\newcommand{\eqr}{\begin{eqnarray}}
\newcommand{\rqe}{\end{eqnarray}}
\newcommand{\eq}{\begin{equation}}
\newcommand{\qe}{\end{equation}}

\newcommand{\eps}{\epsilon}
\newcommand{\pa}{\partial}

\newcommand\preprint[1]{\vspace{-1in}\vtop{\null\hfill
\parbox[t]{1.6in}{\small\sc #1\\\null}}
\vskip .5in\bigskip\normalfont}
\def\KK{{\rm I\kern -.2em  K}}
\def\NN{{\rm I\kern -.16em N}}
\def\RR{{\rm I\kern -.2em  R}}

\def\ZZZ{{\small{\rm Z}\kern -.5em Z}}
\def\QQ{{\rm \kern .25em
             \vrule height1.4ex depth-.12ex width.06em\kern-.31em Q}}
\def\CC{{\rm \kern .25em
             \vrule height1.4ex depth-.12ex width.06em\kern-.31em C}}

\usepackage{epsfig}

\title{\preprint{MIT-CTP-3261\\
hep-th/0204056}
Marginal perturbations of N=4 Yang-Mills as deformations of 
AdS$_5\times S^5$}
\author{}
\date{
Ansar Fayyazuddin$^{1,2}$ 
\thanks{E-mail:~ansar@lns.mit.edu}
and
Subir Mukhopadhyay$^3$
\thanks{E--mail:~subir@physics.umass.edu} \\
\vspace{.8cm}
{\small{\sl 
$^1$ Center for Theoretical Physics,\\ Massachusetts Institute of Technology,\\
Cambridge MA 02139}}\\
\vspace{.8cm}
{\small{\sl 
$^2$Department of Physics,\\Stockholm University,\\
S-106 91 Stockholm, Sweden}}\\
\vspace{.8cm}
{\small{\sl 
$^3$Department of Physics,\\University of Massachusetts,\\
Amherst, MA 01003}} }

\begin{document}
\maketitle

\vspace{0.5cm}

\begin{abstract}
We study constant dilaton supersymmetric solutions of type IIB Supergravity 
with 5-form and 3-form flux with isometry group U(1) $\times$$Z_3$.  Some 
of these solutions correspond to
marginal perturbations of N=4 Yang-Mills.  We find one line of solutions 
in particular of AdS$_5$ fibred over an $S^5$. This line is described by a
single complex parameter.  AdS$_5\times$S$^5$ is obtained when 
this partameter is tuned to zero.
 
\end{abstract}


\thispagestyle{empty}

\newpage

\setcounter{page}{1}

\sect{Introduction}
Much progress has been made in understanding non-perturbative aspects of 
gauge theories in the last few years. A particularly fruitful 
approach to studying gauge theories was suggested by Maldacena who 
conjectured a duality relating quantum field theories to
supergravity (and more generally string theory) in appropriate geometrical
backgrounds.  The conjecture is particularly powerful in the case of conformal 
field theories which are believed to be dual to anti-de 
Sitter spaces.  The reason why the conjecture is most powerful in the CFT case
is that to trust supergravity one normally has to take the 't Hooft coupling 
constant to be large and this leads to problems with defining the dual 
continuum asymptotically free gauge theory.

The greatest success story of Maldacena's conjecture is that of
N=4 Yang-Mills theory 
which is believed to be dual to type IIB supergravity in an AdS$_5\times$S$^5$
background geometry.  In this particular case
physicists have been able to calculate Wilson loop correlators, chiral primary
operator correlation functions etc\footnote{See \cite{review} for an extensive
review and exhaustive references on these topics.}.  
Much is also known about relevant 
perturbations of this gauge theory on the supergravity side\cite{review}.

In this paper we look for supergravity solutions dual to marginal 
deformations of N=4 Yang-Mills theory.  These deformations were 
discovered in the field theory context by Leigh and Strassler
in a beautiful paper\cite{LS}.  In \cite{LS} the authors developed 
non-perturbative methods to establish the existence of
exactly marginal operators which preserve at least N=1 supersymmetry.  

This paper is structured as follows.  In section 2 we briefly describe the 
exactly marginal superpotentials of Leigh and Strassler.  In section 3 we 
describe how to incorporate some of the isometries of the field theory
in the Killing spinor.  In section 4 we solve for the fields of IIB 
supergravity by solving the Killing spinor equations.
In section 5 we impose source equations and Bianchi identities on the
supersymmetric field configurations of section 4.
In section 6 we describe a set of solutions dual to a complex
line of conformal field theories.  Finally, in section 7 we close with some
comments.

\section{Leigh-Strassler deformations of N=4 Yang-Mills theory}

Using field theory techniques, marginal perturbations of ${\cal N}=4$
supersymmetric gauge theory were analyzed by Leigh and Strassler 
in an elegant paper\cite{LS}. 
The essential idea of the analysis is as follows.
Using the fact that the $\beta$ functions \cite{sv1,sv2} of the
couplings depend only on the anomalous dimensions and gauge group
representation of the chiral
fields, one can write down the equations for a
fixed point as a constraint on the anomalous dimensions.
These equations define a hypersurface in the space of couplings.
In general the number of equations
equals the dimension of the space of couplings and thus
the solutions are isolated points in the space of couplings.
However, in the presence of additional symmetries
the number of independent anomalous dimensions gets reduced and
gives rise to a smaller number of constraints resulting in a 
larger subspace of conformal field theories.  Moving in this subspace
corresponds to marginal deformations of the field theory.  
These techniques enable one to obtain 
exactly marginal perturbations of ${\cal N}=4$ theory.

For the sake of completeness let us briefly review the relevant part 
of the analysis of Leigh and Strassler\cite{LS}. 
We start with ${\cal N}=4$ supersymmetric Yang-Mills theory with 
gauge group $SU(n)$ , $n>2$ written in a ${\cal N}=1$ 
language, which corresponds to a choice of a complex structure
in the space of scalars.
The action consists of one vector and three chiral multiplets with a 
superpotential,
${\cal W} = g Tr([\Phi^1,\Phi^2]\Phi^3)$ where $g$ 
is the gauge coupling. In this form only
a $SU(3)\times U(1) \in SU(4)$ part of the R-symmetry is manifest 
where $U(1)$ acts as a common
phase on the chiral fields and $SU(3)$ is the unitary rotation of
the 3 chiral fields.

In order to study marginal perturbations the superpotential 
is perturbed by special additional
terms which generically break the R-symmetry down to $U(1)$ and 
reduce the supersymmetry to ${\cal N}=1$.
The general form for the perturbed superpotential considered in \cite{LS} is
\begin{equation}\label{sup-pot}
{\cal W} = \lambda_1 Tr([\Phi^1,\Phi^2]\Phi^3)  +
           \lambda_2 Tr(\{\Phi^1,\Phi^2\}\Phi^3)  +
           \lambda_3 \sum_{i=1}^3 \Phi_i^3 .
\end{equation}
For generic couplings in (\ref{sup-pot}), 
the R-symmetry is reduced to $U(1)\times Z_3\times Z_3$. 
The first $Z_3$ acts as the group of cyclic permutations on the 
$\Phi_i$'s which  
implies that the anomalous dimensions of
all three chiral fields are equal. 
The second $Z_3$
acts as follows: $(\Phi^1, \Phi^2, \Phi^3) \rightarrow (\omega\Phi^1, 
\omega^2\Phi^2,\omega^3\Phi^3)$, with $\omega^3=1$, it prevents the mixing
of different fields \cite{review}.

The general expressions of the $\beta$ functions for the 
couplings in the superpotential
and for the gauge coupling is obtained explicitly \cite{LS}
as
\begin{eqnarray}
\beta_{\lambda_i} &=& 
\lambda_i(\mu)(-d_w + \sum_{i=1}^3[d_i + \frac{1}{2}\gamma_i] ) ,
\nonumber\\
\beta_g &=& -f(g[\mu])\left( [3C_2(G) - T_i] +\sum_i T(R_i)\gamma_i \right).
\end{eqnarray}
Where $d_w$, $d_i$ are canonical dimensions of the superpotential
and the field $\phi_i$, $\gamma_i$ is the anomalous dimension 
of the chiral fields $\phi_i$. $C_2(G)$ and $T_i$ are the quadratic Casimirs 
of the adjoint representation and the representation of $\Phi_i$ respectively.
$f(g)$ is a function of the gauge coupling which can have a pole
at large $g$.

In the present case (\ref{sup-pot}), due to the $Z_3$ symmetry 
the $\beta$-functions are
reduced to the common anomalous dimension 
$\gamma_i = \gamma(g,\lambda_1, \lambda_2, \lambda_3)$
with scale dependent multiplicative prefactors.
As a result vanishing of the 
$\beta$ functions implies a single equation
\begin{equation}
\gamma(g,\{\lambda_i\}) = 0 ,
\end{equation}
which corresponds to a complex codimension one hypersurface in the
four dimensional space of couplings.
The origin in $\{\lambda_i\}$ is ${\cal N}=4$ supersymmetric 
Yang-Mills theory. A generic 
point, however, corresponds to an ${\cal N}=1$ superconformal theory.

As a consequence of the AdS/CFT correspondence 
there must exist a family of gravity duals corresponding to
the  margnially perturbed conformal theories
which are continuously connected to AdS$_5\times$S$_5$ with isometry
same as that of the superconformal group.
In the following sections we will construct gravity duals of marginal 
perturbations of N=4 Yang-Mills theory.

\section{Constraints on Killing spinors with U(1)$\times$Z$_3$ 
isometry}

As indicated in the previous section, for generic values of the coupling 
constants, $\lambda_i$, the global symmetry is broken from SO(6) down to 
U(1)$\times$Z$_3\times$Z$_3$\footnote{In all that follows we will ignore the
second Z$_3$ simply because we donot understand how to incorporate it without
requiring more symmetry than we want.  Our
explicit solutions do not respect it either.  We are presently investigating
Killing spinors which are invariant under U(1)$\times$SU(3), these would
be invariant under both the Z$_3$s.}.  When using N=1 superspace, 
the maximal manifest
global symmetry is U(1)$\times$SU(3) where the SU(3) rotates the 
three chiral multiplets into each other while the U(1) multiplies them all by
the same phase.  To have a particular N=1 description of the system one
picks a complex structure on the scalars so that they can be organized into
chiral multiplets.  The SU(3) acts in a way which preserves that complex 
structure.  

N=4 Yang-Mills in the AdS/CFT description is realized on the supergravity
side by the near-horizon geometry of D3-branes spanning the directions $0123$.
The transverse directions $456789$ are in one-to-one correspondence with the
scalars of the Yang-Mills theory.  To give an N=1 description we define a
complex structure on the transverse coordinates as follows:
\eq
z^m = x^{m+3}+ix^{m+6}
\qe
where the label $m$ takes values in $1,2,3$.  Now the $z^m$ form a triplet
under the SU(3) subgroup of SO(6), and they have the same charge under the
U(1).

Suppose we have a Killing spinor $\eta$ for a geometry dual to a marginal
deformation.  Then clearly a U(1)$\times$Z$_3$ transform of $\eta$ must
also be a Killing spinor.  While generic elements of SU(3) would transform
$\eta$ into spinors which do not satisfy the Killing spinor equations.  To
investigate the constraints imposed by this simple requirement consider the
generators of SO(6) in the $4\oplus{\overline 4}$ representation:
\eq
J_{ij} = i/2[\hat{\gamma}_i,\hat{\gamma}_j], i,j=4,...,9 
\qe
where $\hat{\gamma}_i$ are flat space $\gamma$-matrices:
\eq
\{\hat{\gamma}_i,\hat{\gamma}_j\} =2\delta_{ij}.
\qe 
Now consider defining $\gamma$-matrices with holomorphic indices corresponding
to the complex structure defined above:
\eqr
\hat{\Gamma}_m & = &\frac{1}{2}(\hat{\gamma}_{m+3}+i\hat{\gamma}_{m+6}),
\nonumber \\
\{\hat{\Gamma}_m,\hat{\Gamma}_n\}&=&0, \\
\{\hat{\Gamma}_m,\hat{\Gamma}_{\overline{n}}\}&=&\delta_{mn},\nonumber\\
\hat{\Gamma}_{\overline{n}}&=&(\hat{\Gamma}_n)^{\dagger}.\nonumber
\rqe
Now consider the following commuting subset of the generators $J_{ij}$:
\eqr
Q &=&\frac{1}{3} (\hat{\Gamma}_{\overline{1}}\hat{\Gamma}_1
+\hat{\Gamma}_{\overline{2}}\hat{\Gamma}_2
+\hat{\Gamma}_{\overline{3}}\hat{\Gamma}_3) -1,\nonumber\\
H_1 &=& \hat{\Gamma}_{\overline{1}}\hat{\Gamma}_1
-\hat{\Gamma}_{\overline{2}}\hat{\Gamma}_2, \\
H_2 &=& \frac{1}{2}(\hat{\Gamma}_{\overline{1}}\hat{\Gamma}_1
+\hat{\Gamma}_{\overline{2}}\hat{\Gamma}_2
-2\hat{\Gamma}_{\overline{3}}\hat{\Gamma}_3).\nonumber
\rqe
We will take $Q$ to generate the U(1) and $H_i$ to be the two Cartan subalgebra
generators of SU(3).  We now decompose the $4\oplus\overline{4}$ representation
 of
the spinors of SO(6) into eigenstates of $Q$.  There are $4$ distinct
eigenvalues: $-1,-1/3,1/3,1$, the subspaces corresponding to these eigenvalues
are given in terms of states built on the spinor $\chi$ which satisfies:
\eq
\hat\Gamma_m\chi = 0.
\qe
The subspaces are:
\eqr
\epsilon_{-1} &=& a^0\chi \nonumber \\
\epsilon_{-1/3}&=& a^i\hat\Gamma_{\overline{i}}\chi\\
\epsilon_{1/3}&=& b^{ij}\hat\Gamma_{\overline{i}}\hat\Gamma_{\overline{j}}\chi
\nonumber\\
\epsilon_{1} &=& b^0\hat\Gamma_{\overline{1}}\hat\Gamma_{\overline{2}}
\hat\Gamma_{\overline{3}}\chi \nonumber.
\rqe
The $a,b$ coefficients are complex numbers.
The subspaces $\epsilon_{1},\epsilon_{-1/3}$ together span the $4$ of SO(6) 
while the remaining two span the $\overline{4}$.  Under SU(3) 
$\epsilon_{\pm 1}$ are singlets, $\epsilon_{-1/3}$ is in the $3$ and 
$\epsilon_{1/3}$ in the $\overline{3}$.

Now we must find an $\eta$ which picks up a constant phase under the U(1), 
but is not invariant under SU(3) except under its discrete Z$_3$ 
subgroup of cyclic permutations. 
Since $\epsilon_{\pm 1}$ are SU(3) singlets neither one of them is a 
possibility.  Instead we must pick a spinor either from $\epsilon_{-1/3}$
or $\epsilon_{1/3}$.  Since either one will do (upto a redefinition of 
complex structure $z^{m}\leftrightarrow z^{\overline{m}}$ they are the 
same) we pick the following combination from $\epsilon_{-1/3}$:
\eq
\eta = \sum_{m=1}^{3}\hat\Gamma_{\overline{m}}\chi.\label{killings}
\qe
Clearly $\eta$ is symmetric under S$_3$ (which contains Z$_3$) and picks up 
a phase under the U(1),
while under generic SU(3) rotations $\eta$ is not invariant.  Note that since
$\eta$ is a chiral spinor in 10d it satisfies the following chiral property:
\eq
\hat{\gamma}^{0123}\eta = i\eta.
\qe

\section{Solving the Killing spinor equations}
In this section we solve the Killing spinor equations assuming that the
dilaton is constant.  The reason why we hold the dilaton constant is that
the dilaton is related to the Yang-Mills coupling constant and should not 
run because of conformal symmetry.

The variation of the dilatino and gravitino under supersymmetry with the
dilaton held constant are as follows\cite{js}:
\eqr
\delta\lambda &=& -\frac{i}{24}\gamma^{ijk}\eta G_{ijk} \nonumber\\
\delta\psi_i &=& D_{i}\eta +\frac{i}{480}\gamma^{i_1i_2i_3i_4i_5}\gamma_i
\eta F_{i_1i_2i_3i_4i_5} + \frac{1}{96}(\gamma_i^{jkl}G_{jkl}-9\gamma^{jk}
G_{ijk})\eta^*.\label{killeq}
\rqe
We take the following metric ansatz:
\eq
ds^2= \Omega^{2}\eta_{\mu\nu}dx^\mu dx^\nu + 2g_{m\overline{n}}dz^{m}
dz^{\overline{n}},
\qe
and assume that $F_5 = {\cal F}_5 + *{\cal F}_5$, with\footnote{We take this 
ansatz so as to preserve SO(1,3) invariance in the $0123$ directions.} 
\eq
{\cal F}_5 ={\cal F}_{0123m}dx^0\wedge dx^1\wedge dx^2\wedge dx^3\wedge dz^m
+ {\cal F}_{0123\overline{n}}dx^0\wedge dx^1\wedge dx^2\wedge dx^3\wedge 
dz^{\overline{n}}.
\qe
Using the explicit form of the Killing spinor $\eta$ given in (\ref{killings})
one finds that the equations (\ref{killeq}) are satisfied if the following
equations hold:
\eqr
F_5 &=& \frac{1}{4}\partial_m\Omega^4dx^0\wedge dx^1\wedge dx^2\wedge 
dx^3\wedge dz^m
+ \frac{1}{4}\partial_{\overline{m}}\Omega^4dx^0\wedge dx^1\wedge dx^2\wedge 
dx^3\wedge dz^{\overline{m}}\nonumber \\
&-&\frac{1}{12}\partial_n\ln\Omega^4\sqrt{\det g}
\hat{\epsilon}_{mpq\overline{rsl}}g^{n\overline{l}}
dz^m\wedge dz^p\wedge dz^q\wedge dz^{\overline{r}}\wedge dz^{\overline{s}}
\nonumber \\
&-&\frac{1}{12}\partial_{\overline{n}}\ln\Omega^4\sqrt{\det g}
\hat{\epsilon}_{\overline{mpq}rsl}g^{l\overline{n}}
dz^{\overline{m}}\wedge dz^{\overline{p}}\wedge dz^{\overline{q}}\wedge 
dz^{r}\wedge dz^{s}
\nonumber \\
G &=& Kdz^{1}\wedge dz^{2}\wedge dz^3 \label{susy}\\
\partial_m(\Omega^{2}g_{n{\overline{p}}}) &=& 
\partial_n(\Omega^{2}g_{m{\overline{p}}}).\nonumber
\rqe 
Here $\hat{\epsilon}_{\overline{mpq}rsl}$ is a completely
anti-symmetric symbol with $\epsilon_{123{\overline{123}}}=-1$.

It is convenient to define the rescaled metric $\hat{g}_{m\overline{n}}=
\Omega^{2}g_{m\overline{n}}$.
This re-scaled metric is Kahler according to (\ref{susy}).  
$\hat{g}$ satisfies an additional constraint which comes from 
the differential equation for the spinor $\chi$ imposed by the equations
in (\ref{killeq}).
This differential equation can be expressed in terms of a function 
$f$ defined by $\chi = f\chi_0$, where $\chi_0$ is a constant spinor:
\eqr
0 &=& \partial_{\overline{m}} \ln f - \frac{1}{2}\partial_{\overline{m}}
\ln\Omega +
\frac{1}{2}\sum_{b}\hat{e}^{p\overline{a}}\hat{e}^{\overline{s}}_{\overline{b}}
\partial_{\overline{m}}\hat{g}_{p\overline{s}} -
\frac{1}{8}\partial_{\overline{m}} \ln\det\hat{g}\\
0 &=& \partial_{m} \ln f - \frac{1}{2}\partial_m
\ln\Omega -
\frac{1}{2}\sum_{b}\hat{e}^{p\overline{a}}\hat{e}^{\overline{s}}_{\overline{b}}
\partial_{m}\hat{g}_{p\overline{s}} +
\frac{1}{8}\partial_m \ln\det\hat{g}\nonumber.
\rqe
The indices $a,b$ are tangent space indices and the $\hat{e}$ are 
vierbein for the $\hat{g}$ metric.  Notice that the index $a$ is 
uncontracted in the two equations above.  Thus the equations must be 
independent of $a$ (this is related to the Z$_3$ symmetry), 
and $\partial_m \ln\det\hat{g}=0$.
These two constraints are solved by constant metrics $\hat g$ 
symmetric under cyclic interchange of indices.  There may be more general
solutions.  
\sect{Bianchi identities and source equations}
Once we have solved the Killing spinor equations we must impose Bianchi
identities as well as the source equations determining the various arbitrary
functions appearing in the above supersymmetric ansatz.  The equations
remaining to be imposed or checked are\cite{js}:
\eqr
dF_5&=& \frac{i}{8}G\wedge\overline{G} + \rho_{local}\nonumber \\
dG&=&0 \label{eom}\\
D^{i_3}G_{i_1i_2i_3} &=& -\frac{2i}{3}F_{i_1i_2i_3i_4i_5}G^{i_3i_4i_5}\nonumber
\rqe
The third equation in (\ref{eom}) is automatically satisfied.  The second
equation gives the constraint:
\eq
\partial_{\overline{m}}K=0.
\qe
In other words $K$ is a holomorphic function of the $z^m$.  
Finally, the first equation imposes:
\eq
\partial_{\overline m}(\sqrt{\det g}g^{n{\overline m}}
\partial_n\ln\Omega^4)+
\partial_m(\sqrt{\det g}g^{m{\overline n}}
\partial_{\overline n}\ln\Omega^4) =-i\frac{1}{8}|K|^2 + \rho_{local}
\qe  
where $\rho_{local}$ collects sources for any D3-branes in the 
problem. Expressing
this last equation in terms of the rescaled metric $\hat{g}$ one can put
it in a somewhat simpler form:
\eq
\partial_{\overline m}(\sqrt{\det \hat{g}}\hat{g}^{n{\overline m}}
\partial_n\Omega^{-4})+
\partial_m(\sqrt{\det \hat{g}}\hat{g}^{m{\overline n}}
\partial_{\overline n}\Omega^{-4}) = -\frac{i}{8}|K|^2 + \rho_{local}
\qe  
Finally using the fact that $\hat{g}$ is Kahler one can further simplify the
equation to:
\eq
2\sqrt{\det\hat{g}}\hat{g}^{m\overline{n}}\partial_m\partial_{\overline{n}}
\Omega^{-4} = -\frac{i}{8}|K|^2 + \rho_{local} \label{main}
\qe
This completes the description of a general supersymmetric solution with
U(1)$\times$Z$_3$ isometry. 

\section{A one-parameter line of solutions}
In this section we discuss a class of solutions describing a one-parameter
family of fibered AdS$_5$ spaces which satisfy the above equations.  Consider
the simplest case where the metric:
\eq
\hat{g}_{m\overline{n}}= \frac{1}{2}\delta_{m\overline{n}}.
\qe
This metric satisfies all the constraints.  The only equation we need to solve
is (\ref{main}), which is an ordinary Laplace equation with a source.  If we
have $N$ D3-branes localized at the origin of $(z^1,z^2,z^3)$ then consider 
the following solution:
\eq
\Omega^{-4} = 1 + \frac{4\pi g_sN\alpha '^2}{r^4} + 
\frac{1}{12}|F(z^1+z^2+z^3)|^2.
\qe
where $r^2 = \sum_{m=1}^{3}|z^m|^2$.  We are assuming an asymptotically flat
solution, hence $F$ must vanish at large $r$.  
Notice that $F$ is a holomorphic function of a 
single variable: $z^1+z^2+z^3$.  This function satisfies (\ref{main}) 
with $K=F'$.  

We will consider a particular form for $F$ so that it survives in the 
near-horizon limit without dominating over the term describing the localized
D3-branes.  Consider then:
\eq
F = -\frac{d\alpha '}{2(z^1+z^2+z^3)^2}.
\qe
We can now take the near-horizon limit of \cite{maldacena}:
$\alpha '\rightarrow 0$ while 
keeping $w^m = z^m/\alpha '$ 
fixed, we also define $u= r/\alpha '$ and $w^m=uf_m(y_i)$, where the $y_i$ are
coordinates on S$^5$ and $\sum_{m=1}^{3}|f_m|^2=1$.  The metric in these new
coordinates is given by
\eq
ds^2 = \alpha '[(u^2\Lambda^2\eta_{\mu\nu}dx^\mu dx^\nu +
\frac{du^2}{\Lambda^2u^2})+\Lambda^{-2}d\Omega_{S^5}^2]
\qe
where $\Omega^2 \equiv\alpha '\Lambda^2u^2$ and $d\Omega_{S^5}^2$ is the 
metric on S$^5$.
The metric is an AdS$_5$ space fibered over a transverse space which is
conformal to an S$^5$.  The radius of curvature at fixed S$^5$ coordinates is:
\eq
R^2_{AdS}=\Lambda^{-2}=\sqrt{4\pi g_sN +\frac{1}{48}
\frac{|d|^2}{|f_1+f_2+f_3|^4}}.
\qe

The 3-form $G$ can be written as:
\eqr
G &=& \alpha '\frac{d}{(z^1+z^2+z^3)^3}dz^1\wedge dz^2\wedge dz^3 \nonumber \\
&=& \alpha '\frac{d}{(w^1+w^2+w^3)^3}dw^1\wedge dw^2\wedge dw^3 \\
&=& \alpha '\frac{d}{(f_1+f_2+f_3)^3}
[df_1\wedge df_2\wedge df_3 + \frac{du}{u}\wedge(f_1df_2\wedge df_3
-f_2df_1\wedge df_3+f_3df_1\wedge df_2)].\nonumber
\rqe 

Both the metric and the $G$-field has an isometry group 
$SO(4,2)\times U(1)\times Z_3$ where the $Z_3$ acts as cyclic permutation
of the labels of $z^m$. 
The symmetry of the solution under the last two factors is obvious.
In order to check the invariance under SO(4,2), recall that it 
consists of the following generators: the generators of Poincare 
transformation acting on $x$'s, scale transformations which act
as $(x,u) \rightarrow (cx , c^{-1}u)$ and special conformal transformations 
which act on both $x$ and $u$.
The isometry under the Poincare and scale transformations
is obvious.
The killing vector for the special conformal transformations
in our coordinates is given by
\begin{equation}
\xi = \eps^\sigma [ -\frac{1}{\Lambda^4 u^2}\pa_\sigma + x^2\pa_\sigma 
- 2x_\sigma(u\pa_u - x^\lambda\pa_\lambda) ]
\end{equation}
A straightforward evaluation of the Lie derivative on the metric shows it is
invariant while for the $G$-field it is sufficient to check that the 
Lie derivative on the 1-form $\frac{du}{u}$ vanishes.  

Notice that this solution is a one complex parameter family of solutions
labeled by $d$.  When $d$ vanishes the space becomes AdS$_5\times$S$^5$ which
is dual to N=4 Yang-Mills theory.  Thus this solution represents a supergravity
dual of a continuous conformal deformation of N=4 Yang-Mills theory.  It would
be interesting to understand its relation to the Leigh-Strassler deformations
and to try to recover the remaining direction of
marginal perturbations predicted in\cite{LS}.  

\sect{Conclusions}
In this paper we studied new supersymmetry preserving solutions of type 
IIB supergravity with a constant dilaton.  Our aim was to discover backgrounds
dual to certain marginal perturbations discovered by Leigh and Strassler
\cite{LS} of N=4 Yang-Mills theory.

Under assumptions of preserving certain isometries we were able to solve the
Killing spinor equations.  The solutions of the Killing spinor equations were
given in terms of a constrained Kahler metric, a holomorphic 3-form
and a function $\Omega$.  These were related to each other by a source
equation.  

N=4 Yang-Mills has a single marginal coupling: the Yang-Mills coupling.  In
this paper we found a solution dual to a gauge theory preserving N=1 
supersymmetry with two marginal couplings.  
These form a new continuous line of supersymmetry preserving solutions 
parameterized by two complex parameters: $\tau=C_0+i/g_s$ and $d$ whose
gauge theory interpretation we defer for the moment.  For a single value 
of the parameter ($d=0$) one recovers
the dual of N=4 Yang-Mills theory.  At other values there are new 
gauge theory duals.

There are a number of open questions.
Although we set out to find the duals of Leigh-Strassler marginal deformations
we have not established the correspondence by relating our $d$ to the 
$\lambda_i$ couplings of the superpotential. 
According to Leigh and Strassler the dimensionality of the space of marginal
couplings is 3.  It would be very interesting to find the remaining marginal
direction.  We hope that more general solutions to our equations will yield
the full space of theories dual to the Leigh-Strassler deformations.  
In particular there are more general ansatze for constant metrics which
we have not investigated and which may yield new marginal directions.  We 
leave this to future work.  Another question is related to the second Z$_3$ 
symmetry group which is not preserved by our solutions.  We have no
understanding of why this symmetry is not present in our solutions.  
One possibility is that the supergravity solution is hinting
at a discrete anomaly in the field theory 
which prevents the Z$_3$ from being realized.  
By acting on our solution with
this Z$_3$ one can generate new solutions, however, the resulting solutions
are not invariant under the first Z$_3$ (the group of cyclic permutations).
  
{\bf Note Added}: As we were completing the paper for submission we received
\cite{sr,oa} which discuss marginal perturbations of 
theories arising from orbifolds of N=4 Yang-Mills theory.    

\section{Acknowledgements}
AF would like to thank Kazutoshi Ohta for discussions and Marcus Quandt for
a helpful discussion. SM would like to thank David Kastor for useful discussions
. 
AF is supported by a grant from Vetenskapsr{\aa}det (Sweden) and in part
by funds provided by
the U.S. Department of Energy (D.O.E.) under cooperative research agreement
\#DF-FC02-94ER40818. SM is supported by NSF Grant PHY-9801875.


\begin{thebibliography}{77}
\bibitem{maldacena}
J.~Maldacena,
``The large $N$ limit of superconformal field theories and supergravity,''
Adv.\ Theor.\ Math.\ Phys.\  {\bf 2}, 231 (1998)
[Int.\ J.\ Theor.\ Phys.\  {\bf 38}, 1113 (1998)]
[arXiv:hep-th/9711200].


\bibitem{LS}
R.~G.~Leigh and M.~J.~Strassler,
``Exactly marginal operators and duality in four-dimensional N=1 supersymmetric
 gauge theory,''
Nucl.\ Phys.\ B {\bf 447}, 95 (1995)
[arXiv:hep-th/9503121].


\bibitem{sv1}
M.~A.~Shifman and A.~I.~Vainshtein,
``Solution Of The Anomaly Puzzle In Susy Gauge Theories And The Wilson Operator
 Expansion,''
Nucl.\ Phys.\ B {\bf 277}, 456 (1986)
[Sov.\ Phys.\ JETP {\bf 64}, 428 (1986)].

\bibitem{sv2}
M.~A.~Shifman and A.~I.~Vainshtein,
``On holomorphic dependence and infrared effects in 
supersymmetric gauge theories,''
Nucl.\ Phys.\ B {\bf 359}, 571 (1991).

\bibitem{review}
O.~Aharony, S.~S.~Gubser, J.~Maldacena, H.~Ooguri and Y.~Oz,
``Large N field theories, string theory and gravity,''
Phys.\ Rept.\  {\bf 323}, 183 (2000)
[arXiv:hep-th/9905111].

\bibitem{js}
J.~H.~Schwarz,
``Covariant Field Equations Of Chiral N=2 D = 10 Supergravity,''
Nucl.\ Phys.\ B {\bf 226}, 269 (1983).

\bibitem{sr}S. S. Razamat,
''Marginal Deformations of N=4 SYM and of its Supersymmetric 
Orbifold Descendants'', hep-th/0204043.

\bibitem{oa} O. Aharony and S. S. Razamat, 
``Exactly Marginal Deformations of N=4 SYM and of 
its Supersymmetric Orbifold Descendants'', hep-th/0204045.

\end{thebibliography}
\end{document}